\documentclass{article}

\usepackage{url}
\usepackage{amsmath}
\usepackage{amssymb}
\usepackage{amsthm}
\usepackage[all]{xy}
\usepackage{amssymb}
\usepackage{verbatim}
\usepackage{graphicx}

\newcommand{\half}{\frac{1}{2}}

\title{The $p^7$ term in the new Expansion for $\lambda_2(p)$ of the Monomer-Dimer Problem}
\author{Paul Federbush \thanks{Department of Mathematics, University of Michigan, Ann Arbor,
 MI 48109-1043, \emph{email}:pfed@umich.edu} }
\date{}

\begin{document}
\maketitle

\begin{abstract}
In a recent paper S. Friedland and the author presented a formal expression for $\lambda_d(p)$ of the monomer-dimer problem in $d$ dimensions involving a power series in $p$. We there presented the result of computations for the terms in the power series through the sixth power. I herein present the result for the seventh power term, in $d=2$. An interesting feature of the new computation is that it would have been impossible without a new algorithm for computation of the Tutte polynomial, by Bjorklund, Husfeldt, Kaski, and Koivisto.
\end{abstract}

In \cite{ctaed} the author presented a formal expansion for $\lambda_d$ of the dimer problem
\begin{align}
\lambda_d \sim \half\ln(2d)- \half+\frac{1}{8}\frac{1}{d} + \frac{5}{96}\frac{1}{d^2}+\frac{5}{64}\frac{1}{d^3}+\cdots. \label{eq:1}
\end{align}
working with Shmuel Friedland this work was extended to the monomer-dimer problem, yielding a formal expansion
\begin{align}
\lambda_d(p) \sim \half(p\ln(2d) - p \ln(p) - 2(1-p)\ln(1-p) -p)+\sum_{k=1}^\infty \frac{c_k(p)}{d^k}\label{eq:2},
\end{align}
with
\begin{align}
c_1(p)&= \frac{1}{8} p^2\\
c_2(p)&= (2p^3 + 3p^4)/96\\
c_3(p)&= (-5p^4+12p^5+8p^6)/192
\end{align}
We have come to believe that it is better to organize the expansion in (\ref{eq:2}) as a power series in $p$. We thus write
\begin{align}
\lambda_d(p)\sim\half(p\ln(2d)-p\ln(p) - 2(1-p)\ln(1-p) -p) + \sum_{k=2}^\infty a_k(d)p^k\label{eq:6}
\end{align}
with
\begin{align}
a_2(p)&= \frac{1}{8} \frac{1}{d}\\
a_3(p)&= \frac{1}{48} \frac{1}{d^2}\\
a_4(p)&= \frac{1}{32}\frac{1}{d^2} - \frac{5}{192}\frac{1}{d^3}\\
a_5(p)&= \frac{1}{16} \frac{1}{d^3} - \frac{39}{640} \frac{1}{d^4}\\
a_6(d)&= \frac{1}{24}\frac{1}{d^3} - \frac{1}{32}\frac{1}{d^4} - \frac{19}{1920}\frac{1}{d^5}
\end{align}

Each of the three expansions is computed as far as it can be from knowledge of the kernels $\bar{J_i}$ computed to construct series (\ref{eq:1}). These $\bar{J_i}$ are given in (31)--~(36) of \cite{ctaed}, and repeated in (5.25)--(5.30) of \cite{aaeri}. The $\bar{J_i}$ were obtained by lengthy computations, by machine for $i=5,6$. Once the $\bar{J_i}$ are known each of the three series above is given by an easy computation (a short run of a short Maple program).

When I first constructed these series I viewed them as asymptotic, series (\ref{eq:1}) valid in the limit $d\to\infty$. And series (\ref{eq:6}) for $p$ small enough. But we now believe series (\ref{eq:6}) converges for all physical values! I.e. $d=1,2,\cdots, 0\leq p \leq 1$. As was pointed out in \cite{aaeri}, (6)--(11) is true for $d=1$; and for $d=2$, (6)--(11) is strongly consistent with a numerical study of $\lambda_2(p)$.

For $d=2$, the expression from (6)--(11), and the computation of the present paper, is 
( first  $\overline{J}_7=  \frac{299}{14336}$  and therefrom )
\begin{align}
\lambda_2(p)&\sim \half (p \ln(4) - p\ln p - 2(1-p) \ln(1-p) -p) \nonumber\\
&+ \frac{1}{2^4} p^2 + \frac{1}{2^6\cdot 3} p^3 + \frac{7}{2^9\cdot 3} p^4 + \frac{41}{2^{11}\cdot 5} p^5 \\
&+\frac{181}{2^{12}\cdot 3\cdot 5} p^6 +\frac{757}{2^{14}\cdot 3\cdot 7} p^7\nonumber
\end{align}

The computation of the $p^7$ term involved 1000+ hours' running of a Maple program on my p.c. The computation of the contribution of each perturbation term was done in integer arithmetic, but to avoid ridiculous fractions at intermediate steps, the sum of contributions was done in floating point. The interesting feature of the computation involved the computation of the expression $\psi'_c$, see eq. (24) of \cite{ctaed} or eq. (5.21) of \cite{aaeri}. The overlap pattern of the dimers determines a graph with 7 vertices each representing a dimer, and edges between verticies whose dimers overlap. $\psi_c'$ as defined in \cite{ascice}, when computed from its definition, requires $2^{21}$ computer subroutines, $2^{21}$ being the number of possible subsets of the edges. I learned from Richard Kenyon that $\psi_c'$ is, up to sign, $T_G(1,0)$, $T_G$ the Tutte polynomial of the graph. And in \cite{tutte} an algorithm is given which has an exponential in the number of vertices, instead of the number of edges, determining its computation time (to compute $T_G$).\footnote{In fact I implement the "baseline algorithm" of \cite{tutte}, Section 3.1 therein, but using the basic zeta and Moebius transforms rather than their fast versions} Use of this algorithm made the computation possible.

\section*{Acknowledgements}
Thanks to Richard Kenyon for informing me about the Tutte polynomials. I would also like to thank Bennet Fauber for doing some of the computer runs for me.

\nocite{cfelm}
\bibliographystyle{plain}
\bibliography{bib}
@article{ctaed,
    author = "Federbush, P.",
    title = "Computation of Terms in the Asymptotic Expansion of Dimer $\lambda_d$ for High Dimensions",
    journal= "Physics Letters A",
    year = "2009",
    volume = "374",
    pages = "131---133"
}

@article{aaeri,
    author = "Federbush, P. and Friedland, S.",
    title= "An Asymptotic Expansion and Recursive Inequalities for the Monomer-Dimer Problem",
    journal= "Journal of Statistical Physics",
    volume = "143", 
    year = "2011",
    pages="p.~306"
}

@unpublished{cfelm,
    author = "Federbush, P.",
    title= "Convergence of the Formal Expansion for $\lambda_d(p)$ of the Monomer-Dimer Problem for small $p$",
    note=" arXiv: math-ph/1101.4591"
}

@inproceedings{tutte,
	author="Bjorklund, A. and Husfeldt, T. and Kaski, P. and Koivisto, M.",
	pages="pp.~677---686",
	year="2008",
	title="Computing the Tutte Polynomial in Exponential Time ",
	booktitle="49th Annual IEE Symposium on Foundations of Computer Science",
	series="FOCS"
}

@incollection{ascice,
	author="Brydges, D. C.",
	title="A Short Course in Cluster Expansions",
	booktitle="Phenomenes Critiques, Systems Aleatoires, Theories de Guage, parts I, II",
	year="1986",
	address="North-Holland, Amsterdam",
	pages="pp. 129---183"
}

\end{document}